\documentclass[journal,draftcls,onecolumn,11pt]{article}
  \usepackage{epsfig}
\begin{document}
\title{Maximum lifetime problem in sensor networks with limited channel capacity}
\author{
Z. Lipi{\'n}ski  \\
        Institute of Mathematics and Informatics \\
        Opole University, Poland
}
\maketitle
\begin{abstract}
In the paper we analyze the maximum lifetime problem
in sensor networks with limited channel capacity
for multipoint-to-multipoint and broadcast data transmission services.
We show, that in order to achieve an optimal data transmission regarding the maximum lifetime problem we cannot allow for any interference of signals.
We propose a new Signal to Interference plus Noise Ratio function
and used is to modify the Shannon-Hartley channel capacity formula.
For the modified channel capacity formula we solve the maximum lifetime problem
in one dimensional regular sensor network $L_N$
for discussed data transmission services.
\end{abstract}
\section{Introduction}
Characteristic feature of sensor networks is that these consist of small electronic
devices with limited power and computational resources.
Most of the energy the nodes of the sensor network possess is being utilized in the process of data transmission.
The energy consumption depends on the size of the network, amount of data transmitted over the network
and also on the presence of noise and interference of signals within the transmission channels.
On the other hand, from the network a long term operating time is expected,
that the functional lifetime and services delivered by the network can be available for a long time.
It seems that, one of the most important problem in sensor networks is to
optimize the energy consumption of each node to extend their operating time and thus the lifetime of the whole network.
We define a sensor network lifetime as the time until the first node of the network runs out of energy,
\cite{Giridhar, Acharya}.
If each node of the network has a battery with an initial energy $E_0$,
then by finding the optimal energy utilization of each node  
we can determine the number of cycles
$N_{{\rm cycles}}=[\frac{E_0}{E_{i'}^{{\rm opt}}}]$
the network can perform its functions until the most overloaded node runs out of its energy $E_{i'}^{{\rm opt}}$.

The energy consumption model in wireless sensor networks can be expressed in terms of
power $P_i$ and the data transmission time $t_i$ of each sensor.
For the optimization problems in which we minimize the energy consumed by each node
%
 \begin{equation} \label{NodeEnergy-Time}
E_i = P_i \; t_i,
 \end{equation}
we must assume that capacities of the transmission channels are limited,
otherwise the minimum energy of each node is reached for $t_i =0$.
The capacity of a noisy channel can be described by the well know Shannon-Hartley formula, \cite{Franceschetti},
 \begin{equation} \label{Shannon-Hartley}
C = B \log (1 + \frac{P}{\textit{N}_o}),\;\;[b/s].
\end{equation}
where $B$ is the channel bandwidth,
$P$ the signal and $\textit{N}_o$ the noise power.
An impact on the channel capacity has not only the level of a noise
but also the interference of signals generated simultaneously by nodes in a given network.
To take into account the interference in wireless network the quotient $\frac{P}{\textit{N}_o}$
in the Shannon-Hartley formula (\ref{Shannon-Hartley}) should be replaced by a Signal to Interference plus Noise Ratio function.
Let us denote by $S_N$ the wireless sensor network built of $N$ nodes.
We assume, that the nodes are located at the points $x_i$ of $d$-dimensional space $R^d$.
By $P(x_i)$ we denote the transmission power of the $i$-th node and
by $\gamma(x_i,x_j)$ the signal gain function between two nodes located at the points $x_i$ and $x_j$.
Because of the loss factor $\gamma(x_i,x_j)$
the strength of signal of the $i$-th node detected by the $j$-th node is equal to $P(x_i) \gamma(x_i,x_j)$.
The Signal to Interference plus Noise Ratio (SINR) function defined in \cite{Gupta},
see also \cite{Franceschetti, Baccelli, Grossglauser}, is given by the formula
 \begin{equation} \label{SINR-Gupta-Kumar}
  s(x_i,x_j,U_{i}) = \frac{ P(x_i) \gamma(x_i,x_j) }{ \textit{N}_o + \sum_{k \in U_{i}} P(x_k) \gamma(x_k,x_j) },
  \end{equation}
where $U_{i}\subset S_N$ is the set of nodes which signals interfere with the signal of given $i$-th node.
By definition $i \notin U_{i}$ for any $i\in S_N$.
Note that, when the nodes use the omnidirectional antennas and their transmission power
is such that the signal is detected by any other node of the network,
then $U_{i}$ is the set of nodes which simultaneously transmit data with the $i$-th node.
The formula (\ref{SINR-Gupta-Kumar}) allows to estimate the conditions under which the transmission from the $i$-th node is successfully
received by the $j$-th node.
In literature a typical threshold $\beta \leq  s(x_i,x_j,U_{i})$ below which
the data cannot be transmitted has value of $\beta \geq 1.1$.
From the formula (\ref{SINR-Gupta-Kumar}) it follows that the parameter $\beta$ limits the number of nodes in the set $U_{i}$.
The modified Shannon-Hartley formula can be written in the form
 \begin{equation} \label{MaxTransRateG-K}
 C(x_i,x_j,U_{i}) = \log (1 + s(x_i,x_j,U_{i})),
 \end{equation}
where $s(x_i,x_j,U_{i})$ is the SINR function (\ref{SINR-Gupta-Kumar}).
In (\ref{MaxTransRateG-K}) the channel bandwidth $B$ we put equal to one.
The formula (\ref{MaxTransRateG-K}) allows to determine the maximum achievable transmission rate between the $i$-th and $j$-th node
in the presence of other transmitters.
If the sensors utilize the point-to-point data transmission in the physical layer
then the amount of data $q_{i,j}$ transmitted between $i$-th and $j$-th node is a product
of the transmission rate $c_{i,j}$ and transmission time $t_{i,j}$ of the $i$-th node
 \begin{equation} \label{qij-cij-tij}
            q_{i,j}  = c_{i,j}\; t_{i,j}.
 \end{equation}
In optimization problems discussed in this paper the transmission rate matrix elements $c_{i,j}$ used as
parameters satisfy the inequality
 \begin{equation} \label{TransRate-Inequality}
 0 \leq c_{i,j} \leq C(x_i,x_j,U_{i}),
 \end{equation}
where $C(x_i,x_j,U_{i})$ is given by (\ref{MaxTransRateG-K}).

Another model of energy consumption in wireless networks can be described in
terms of data transmission cost energy matrix $E_{i,j}$ and the data flow matrix $q_{i,j}$,
\cite{Giridhar, Chang}.
Elements of the matrix $E_{i,j}$ define
the energy required to transmit one unit of data between two nodes. 
The energy necessary to transmit the amount of data $q_{i,j}$ from the $i$-th node to the $j$-th node
is a product $E_{i,j} q_{i,j}$ and
the total the energy $E_i$ consumed by the $i$-th node to transmit all of its data
is given by 
 \begin{equation} \label{NodeEnergy-q}
E_i = \sum_{j} q_{i,j} E_{i,j}.
 \end{equation}
In the paper we discuss the relation between these two energy consumption models (\ref{NodeEnergy-Time}) and (\ref{NodeEnergy-q}).
We show, that for a modified SINR function (\ref{SINR-Gupta-Kumar}), such that
the signal of the transmitting node depends on the distance to the receiver,
the solution of the maximum lifetime problem for both energy consumption models coincide.
We also show, that for a sufficiently large network the solution of the maximum lifetime problem
with the SINR function given by (\ref{SINR-Gupta-Kumar}) can be reduced to the solution of the problem
discussed in the papers \cite{Giridhar, Chang, Lipinski1, Lipinski2},
where the energy consumption model (\ref{NodeEnergy-q}) was utilized.

In this paper we consider two types of the maximum lifetime problems in sensor networks
with limited channel capacity and in the presence of noise and interference of signals.
The first problem is related to the optimization of the data transmission
from a given set of sensors to the set of data collectors in $S_N$.
This type of problem we call later on
a multipoint-to-multipoint data transmission service. 
The second problem describes the maximum lifetime problem for the broadcast data transmission.
The objective function of the maximum lifetime problem
for both types of services
has the form
 \begin{equation} \label{ObjectiveFunction}
E_{i'} = \max_{i\in S_N} \{ E_{i} \},
 \end{equation}
where $E_{i}$ is the energy consumed by the $i$-th node of the network $S_N$.
This formula can be utilized for the energy consumption model (\ref{NodeEnergy-Time}) and (\ref{NodeEnergy-q}).
%
In the physical layer the nodes of wireless network can use the point-to-point or the point-to-multipoint
data transmission.
For the point-to-point transmission the sender transmits data to the unique receiver.
For the point-to-multipoint data transmission the transmitter sends parallel the same data to a set of receivers.
An important future of the point-to-multipoint data transmission is the 'wireless multicast advantage' property,
\cite{Wieselthier}.
For such transmission the nodes, which are in the range of the transmitting node, can receive the data
without additional cost of the transmitter.
In this paper we consider the maximum lifetime problems in sensor networks which nodes
use in the physical layer the point-to-point data transmission.

For the multipoint-to-multipoint data transmission service
the minimum of the objective function (\ref{ObjectiveFunction})
is determined under the data flow transmission constraint
 \begin{equation} \label{DataFlowConstraintM2P}
 \sum_{i} q_{i,j} = Q_i + \sum_{j} q_{j,i}.
 \end{equation}
This equation states that the amount of data generated by the $i$-th node $Q_i$
and the amount of data received from other nodes $\sum_{j} q_{j,i}$ must be equal to the amount of data which the node can send $\sum_{i} q_{i,j}$.
By definition the data collectors does not generate, transmit or retransmit any data, i.e.,
for a data collector $k\in S_N$,  $\forall_{j\in S_N}\; q_{k,j}=0$.

For the broadcast data transmission service, when the $k$-th node sends the amount $Q_k$ of data to all other nodes of the network,
the requirement that each node of $S_N$ receives this data
can be written in the form
\begin{equation} \label{DataFlowConstraintBroadcast}
 \forall_{j\in S_N, j\neq k} \;\sum_{i, i\neq j}^{} q_{i,j}^{} = Q_k.
\end{equation}

In this paper we show that the optimal transmission for the maximum lifetime problem with
limited channel capacity
is achieved when there is no interference in the transmission channels.
For models in which the nodes use the omnidirectional antennas and the threshold $\beta$ is equal to zero,
the requirement that there is no interference means that the data is transmitted by the nodes sequentially.
We propose a new point-to-point data transmission model with SINR function in which
the signal power of the transmitting node $P(x_i)$ depends on the distance to the receiver
and has the form $P(x_i)=P_0 \gamma(x_i,x_j)$, where $\gamma(x_i,x_j)$ is the signal gain function.
We show, that for such model
the solutions of the maximum lifetime problem for both energy consumption models (\ref{NodeEnergy-Time}) and (\ref{NodeEnergy-q})
can be written in the same form and thus are equivalent.
We give the solution of the maximum lifetime problem with the new SINR function
in the one dimensional regular sensor network $L_N$ for both
the multipoint-to-multipoint and broadcast data transmission services and
show that these solutions coincide with
the analytical solution of the problems discussed in \cite{Giridhar, Lipinski1, Lipinski2}.
We also show, that when the distance between sensor nodes is sufficiently large,
the solution of the maximum network lifetime problem with Gupta-Kumar SINR function (\ref{SINR-Gupta-Kumar}) coincides
with the solution of the problem written in terms of $E_{i,j}$ and $q_{i,j}$ matrices discussed in
\cite{Giridhar, Lipinski1, Lipinski2}.
\section{Data transmission model with the SINR function}
We assume, that the power of the transmitting signal at the receiver
must have some minimal level $P_{0}$.
In other words, $P_{0}$ is the minimal signal power the receiver node can hear.
This requirement means, that the transmitting node must generate the signal with the strength
 \begin{equation} \label{P2P-SignalPower}
          P_{i,j} = P_{0} \;\gamma_{i,j}^{-1},
 \end{equation}
where $\gamma_{i,j}=\gamma(x_i,x_j)$ is a signal gain function between two nodes located at the points  $x_i$ and $x_j$ of $R^d$.
The energy required to send data between the $i$-th and $j$-th node in the time $t_{i,j}$ equals
 \begin{equation} \label{EnergyGammaTime}
        E_{i,j}= P_0 \gamma_{i,j}^{-1} t_{i,j}.
  \end{equation}
When sensors use in the physical layer the point-to-point data transmission the total energy consumed
by the i-th node is given by the formula
 \begin{equation} \label{NodeEnergySINR}
E_{i}(t)= P_0 \sum_{j\in S_N} \gamma_{i,j}^{-1} t_{i,j}.
  \end{equation}
Because the total operating time of the i-th sensor is given by the sum $\sum_{j\in S_N} t_{i,j}$
over all network nodes, the assumption that sensors use the point-to-point data transmission in the physical layer
is important.
We denote by $U \subset S_N \times S_N$ the set of transmitter-receiver pairs $(i,k), \;i,k\in S_N$
such that for two elements $(i,k)$ and $(i',k')$ of $U$
the nodes $i$ and $i'$ transmit data to the $k$ and $k'$ nodes respectively and their signals interfere.
We assume, that $\;i\neq i'$, $k\neq k'$.
For wireless networks in which the nodes use the omnidirectional antennas and the signal is received in the whole network
$U$ can be defined as a set of node pairs which transmitters transmit data simultaneously, i.e.,
 \begin{equation} \label{Set-U-Model2}
 U = \{ (i,k),(i',k') \in S_N\times S_N | t^{(s)}_{i,k}=t^{(s)}_{i',k'}, \;t^{(e)}_{i,k}=t^{(e)}_{i',k'} \},
\end{equation}
where $t^{(s)}_{i,k}$ and $t^{(e)}_{i,k}$ is the start and the end of transmission time between $i$-th and $k$-th node.
If $(i,k) \in U$ then by $U_{i,k}$ we denote the set
$U_{i,k} = U \setminus \{(i,k)\}.$
We assume, that each node cannot receive data simultaneously from many transmitters
and cannot send data to several receivers, i.e. there is no wireless multicast advantage.
For the transmission model (\ref{P2P-SignalPower})
the maximum achievable transmission rate between the $i$-th and $j$-th nodes can be given by the formula
 \begin{equation} \label{MaxAchievTransRateModel2}
       C(x_i,x_j,U^n_{i,j}) =
       \log (1 + \frac{P_0}{ \textit{N}_o +  P_0 \sum_{(k,m) \in U^n_{i,j}} \gamma(x_k,x_m)^{-1} \gamma(x_k,x_j)^{}}),
 \end{equation}
where $U^n_{i,j}$ is a given set of transmitter-receiver pairs
which signal of the transmitters interfere with the signal of the $i$-th node.

In the next sections we show, that
the solutions of the maximum network lifetime problem with the channel capacity (\ref{MaxAchievTransRateModel2})
for the multipoint-to-multipoint and the broadcast data transmission services
for both energy consumption models (\ref{NodeEnergy-Time}) and (\ref{NodeEnergy-q}) are equivalent.
We show that, the solution of the maximum network lifetime problem with maximum transmission rate (\ref{MaxAchievTransRateModel2})
for both types of services in the one dimensional regular sensor network $L_N$
can be obtained from the analytical solutions of the problems given in
\cite{Giridhar,  Lipinski1, Lipinski2}.
 \section{Multipoint-to-multipoint data transmission service }
The maximum network lifetime problem for the multipoint-to-multipoint data transmission service
with the channel capacity given by (\ref{MaxAchievTransRateModel2})
is defined by the objective function (\ref{ObjectiveFunction})
in which the energy $E_i$ of each node is given by the formula
 \begin{equation} \label{NodeEnergyUnSINR}
 E_{i}(t)= P_0 \sum_{j, n} \gamma_{i,j}^{-1} \; t^n_{i,j},
 \end{equation}
where $t^n_{i,j}$ is the data transmission time between the i-th and j-th nodes in the presence of other
transmitters from the set $U^n_{i,j}$.
The data flow transmission constraint (\ref{DataFlowConstraintM2P}) can be written in the form
 \begin{equation} \label{DataFlowConstraintM2P-SINR} \left\{   \begin{array}{l}
   \sum_{i,n} c^n_{i,j} t^n_{i,j} = Q_i + \sum_{j,n} c^n_{j,i} t^n_{j,i},\\
   0 \leq c^n_{i,j} \leq C^n_{i,j},
 \end{array} \right.  \end{equation}
where $C^n_{i,j}$ 
is the maximum transmission rate between $i$-th and $j$-th nodes calculated from (\ref{MaxAchievTransRateModel2}).
For the maximum network lifetime problem of the multipoint-to-multipoint data transmission service
with limited channel capacity we deduct the following.

%
{\bf Theorem 1}. The optimal data transmission for the maximum network lifetime problem
(\ref{ObjectiveFunction}), (\ref{NodeEnergyUnSINR}), (\ref{DataFlowConstraintM2P-SINR})
is the transmission without interference.

{\it Proof}.
The energy consumed by the i-th node
$E_i(t) = P_0 \sum_{j,n} t^n_{i,j} \gamma_{i,j}^{-1}$
is minimal when the data transmission time $t^n_{i,j}$ is minimal.
Because of the constrain (\ref{DataFlowConstraintM2P-SINR}),
the time parameter $t^n_{i,j}$ reaches the minimal value when $c^n_{i,j}=C^n_{i,j}$
and the capacity $C^n_{i,j}$ of the channel $(i,j)$ is maximal.
The $C^n_{i,j}$ is maximal when the interference factor in
the SINR function $s(x_i,x_j,U^n_{i})$ or $s(x_i,x_j,U^n_{i,j})$
is equal to zero.
It means, that the set $U^n_{i}$ in (\ref{SINR-Gupta-Kumar}) and $U^n_{i,j}$ in (\ref{MaxAchievTransRateModel2}) must be empty
and there is no interference of signals in the network.
$\diamond$

The result of Theorem 1 we use to redefine the maximum network lifetime problem
of the multipoint-to-multipoint data transmission service
to the form
 \begin{equation} \label{MLProblemM2PNoInterferece}
 \left\{   \begin{array}{l}
   \min_{t} \max_{i} \{ E_{i}(t) \},  \\
E_i(t) = P_0 \sum_{j} \gamma_{i,j}^{-1} \; t_{i,j},  \\
c_0 \sum_{i} t_{i,j} = Q_i + c_0 \sum_{j} t_{j,i},\\
t_{i,j}\geq 0,\;0\leq c_0 \leq C_0, \; i,j\in [1,N],
  \end{array} \right.\end{equation}
where by $C_0$ we denoted the maximal transmission rate in (\ref{MaxAchievTransRateModel2}),
achieved when $U^n_{i,j}=\emptyset$, i.e.,
 \begin{equation} \label{MaxAchievTransRateNoInter}
   \forall_{i,j}\;\;    C(x_i,x_j) = \log (1 + \frac{P_0}{\textit{N}_o}) = C_0.
  \end{equation}
In the next lemma we give the solution of the maximum network lifetime problem (\ref{MLProblemM2PNoInterferece})
for a one dimensional, regular sensor network $L_N$ with one data collector
for arbitrary signal gain function of the form
$$ \gamma^{-1}(\bar{a},\bar{\lambda}, x_i,x_j) = \sum_{n=0}^{\infty} \lambda_{n} d(x_i,x_j)^{a_n},$$
where $d(x_i,x_j)$ is Euclidean distance in $R$ and $\sum_{n=0}^{\infty} \lambda_{n} =1$, $\lambda_n\geq 0$, $a_n \geq 1$.
We assume, that sensors of the $L_N$ network are located at the points $x_i=i$ of a line and the data collector
at the point $x_0=0$.
For $d_{i,j}=|i-j|$ any two sensors in $L_N$,  which distance between them is equal to $|i-j|=r$
the signal gain $\gamma_{i,j}$ can be written in the form
\begin{equation} \label{gamma-1[r]}
\gamma^{-1}_{r}(\bar{a},\bar{\lambda}) = \sum_{n=0}^{\infty} \lambda_{n} r^{a_n}, \;\; r \in [1,N],
\end{equation}
where $\sum_{n=0}^{\infty} \lambda_{n} =1$, $\lambda_n\geq 0$ and $a_n \geq 1$.
In the following lemma we give the solution of (\ref{MLProblemM2PNoInterferece})
for any gain function (\ref{gamma-1[r]}).

{\bf Lemma 1}. The solution of the maximum lifetime problem
for the multipoint-to-multipoint data transmission service
in the one dimensional regular sensor network $L_N$, with the channel gain function (\ref{gamma-1[r]})
and the channel capacity (\ref{MaxAchievTransRateModel2})
is given by the following set of data transmission trees
$$T^{1}=T_{1,0}, \;\; T^{i}=\{ T^{i}_{i,0},T^{i}_{i,i-1}\}, \;\;i\in [2,N],$$
and weights 
\begin{equation} \label{qijRecurrenceSol}\left\{   \begin{array}{l} 
t_{1,0}  =  \frac{1}{C_0} Q_{N}
          + \frac{1}{C_0}\sum_{j=2}^{N} Q_{j-1} \prod_{r=j}^{N}(1-\frac{\gamma^{-1}_{1}}{\gamma^{-1}_{r}}), \\
t_{2,0} =\frac{1}{C_0} \frac{\gamma^{-1}_{1}}{\gamma^{-1}_{2}} Q_{1}, \\
t_{i,0} =
    \frac{1}{C_0} \frac{\gamma^{-1}_{1}}{\gamma^{-1}_{i}} (Q_{i-1}
+   \sum_{j=1}^{i-2} Q_{i-j-1} \prod_{r=1}^{j}(1-\frac{\gamma^{-1}_{1}}{\gamma^{-1}_{i-r}})),
 \;\;\;\; i \in [3,N],\\
t_{2,1} =\frac{1}{C_0} Q_{2} - \frac{1}{C_0}\frac{\gamma^{-1}_{1}}{\gamma^{-1}_{2}} Q_{1}, \;\;\;\; N=2,\\
t_{2,1} = \frac{1}{C_0} \sum_{k=2}^{N} Q_{k}
        - \frac{1}{C_0} \sum_{k=2}^{N} \frac{\gamma^{-1}_{1}}{\gamma^{-1}_{k}} Q_{k-1}
        - \frac{1}{C_0} \sum_{k=3}^{N} \frac{\gamma^{-1}_{1}}{\gamma^{-1}_{k}} \sum_{j=1}^{k-2} Q_{k-j-1} \prod_{r=1}^{j} (1-\frac{\gamma^{-1}_{1}}{\gamma^{-1}_{k-r}}),
\;\;\;\; N\geq 3,\\
t_{i,i-1} = \frac{1}{C_0} \sum_{k=i}^{N} Q_{k}
          - \frac{1}{C_0} \sum_{k=i}^{N} \frac{\gamma^{-1}_{1}}{\gamma^{-1}_{k}} ( Q_{k-1}
          + \frac{1}{C_0} \sum_{j=1}^{k-2} Q_{k-j-1} \prod_{r=1}^{j} (1-\frac{\gamma^{-1}_{1}}{\gamma^{-1}_{k-r}})),
\;\;\;\; i \in [3,N],
\end{array} \right.
\end{equation}

{\it Proof}.
It is easy to show that the formulas given in (\ref{MLProblemM2PNoInterferece}) by a simple transformation
$q_{i,j}\rightarrow P_0 t_{i,j}$ and $E_{i,j} \rightarrow \gamma^{-1}_{i,j}$ and $Q_{i} \rightarrow \frac{P_0}{C_0} Q_{i}$
can be obtained from the formulas (8) from \cite{Lipinski1},
and the solution (\ref{qijRecurrenceSol}) can be obtained by these transformation from
the solution (16) given in \cite{Lipinski1}.
Detailed proof that the set of trees and weight polynomials (\ref{qijRecurrenceSol}) in the Lemma 1 is optimal
can be carried out analogously to the proof given in \cite{Lipinski1}. $\diamond$
%
 \section{Broadcast data transmission service}
To define the maximum network lifetime problem for the broadcast data transmission
we represent the sensor network $S_N$ as a directed, weighted graph
$G_N=\{S_N, V, E \},$
in which $S_N$ is the set of graph nodes, $V$ is the set edges and $E$ the set of weights.
Each directed edge $T_{i,j}\in V$ defines a communication link between i-th and j-th node of the network.
To each edge $T_{i,j}$ we assign a weight $E_{i,j}$, which is the cost of transmission of one unit of data between
$i$-th and $j$-th node.
By $U^{({\rm out})}_{i} \subseteq S_N$ we denote a set of the network nodes
to which the $i$-th node can sent the data
$U^{({\rm out})}_{i}=\{ x_j \in S_N | \;\exists \;T_{i,j}\in V \}.$
The set $U^{({\rm out})}_{i}$ defines the maximal transmission range of the $i$-th node.
We assume, that each node of $S_N$ can send data to any other nodes of the network, i.e.,
\begin{equation} \label{TheSetUFullRange}
\forall_{i\in [1,N]}, \;\; U^{({\rm out})}_{i} = S_N.
\end{equation}
We describe  the data flow in the network $S_N$ in terms of spanning trees of the graph $G_N$.
The set of all spanning trees $T^{k,r}$, which begin in the $k$-th node we denote by $V_{k}$.
If the assumption (\ref{TheSetUFullRange}) is satisfied,
then $G_N$ is a complete graph and the number of spanning trees rooted at the k-th node $|V_{k}^{}|$
is equal to $N^{N-2}$, \cite{Berge}.
By $q^{k,r}_{i,j}$ we denote the amount of data which is transmitted along the edge $T^{k,r}_{i,j}$.
The amount of data $q_{i,j}^{k}$ send by the $i$-th node to the $j$-th node along all trees is given by the formula
\begin{equation} \label{qijtkr}
q_{i,j}^{k} = \sum_{r}^{} q^{k,r}_{i,j} T^{k,r}_{i,j}.
\end{equation}
We require, that along each tree the transmitted data is the same.
This means that, for fixed $k$ and $r$, the weights $q^{k,r}_{i,j}$ of the edges $T^{k,r}_{i,j}$ are equal.
We denote these weights by $q^{k}_{r}$, i.e.,
\begin{equation} \label{t-q}
\forall_{i,j} \; q^{k,r}_{i,j} = q^{k}_{r}.
\end{equation}
In the energy consumption model (\ref{NodeEnergy-Time}), when the data is transmitted without interference,
the parameter $q^{k}_{r}$ will be replaced by the parameter $t^{k}_{r}$
which defines the transmission time for the amount $q^{k}_{r}$ of data along the $r$-th tree.
Because we assumed, that all data for the broadcast data transmission service
is transmitted along some set of trees,
we must modify in the channel capacity formula (\ref{MaxAchievTransRateModel2})
the definition of the set $U^{n}_{i,j}$.
We replace it by $U^{k,n,r}_{i,j}$, the set of sensors which signal interferes with
the signal of the $i$-th node when the broadcasted data is transmitted along the tree $T^{k,r}$.
We denote the new channel capacity function by $C(x_i,x_j,U_{i,j}^{k,n,r})$.

The maximum network lifetime problem of the broadcast data transmission service
with the channel capacity (\ref{MaxAchievTransRateModel2}) with the set $U^{k,n,r}_{i,j}$
is defined by the objective function (\ref{ObjectiveFunction}) in which the
energy of each node is given by the formula
\begin{equation} \label{NodeEnergySinrBroadcast}
E_{i}(t^{k})= P_0 \sum_{n,r,j}^{} \gamma_{i,j}^{-1} \;t^{k,n,r}_{i,j} \; T^{k,r}_{i,j},
\end{equation}
where $\gamma_{i,j}^{-1}$ is given by (\ref{gamma-1[r]}).
The data flow transmission constraint (\ref{DataFlowConstraintBroadcast}),
which is the requirement that each node of the network must receive data broadcasted by the k-th node,
can be written in the form
\begin{equation} \label{DataFlowConstraintBroadcastSINR}
\left\{   \begin{array}{l}
\sum_{n,r,i} c^{k,n,r}_{i,j} \;t^{k,n,r}_{i,j} \;T^{k,r}_{i,j} = Q_k,\;\; j\in [1,N],\\
t^{k,n,r}_{i,j}\geq 0, \;0 \leq c^{k,n,r}_{i,j} \leq C^{k,n,r}_{i,j},
\;\;T^{k,r} \in V_{k}^{}, \; r\in [1,N^{N-2}],
\end{array} \right.
\end{equation}
where $C(x_i,x_j,U_{i,j}^{k,n,r}) \equiv C^{k,n,r}_{i,j}$ is given by (\ref{MaxAchievTransRateModel2}).
For the maximum network lifetime problem of the broadcast data transmission service defined by
(\ref{ObjectiveFunction}), (\ref{NodeEnergySinrBroadcast}), (\ref{DataFlowConstraintBroadcastSINR})
we propose a theorem, similar to the Theorem 1 for the multipoint-to-multipoint data transmission
service.

{\bf Theorem 2}. The optimal data transmission for the maximum network lifetime broadcasting problem
(\ref{ObjectiveFunction}), (\ref{NodeEnergySinrBroadcast}), (\ref{DataFlowConstraintBroadcastSINR})
       is the transmission without interference.

{\it Proof}. The proof is analogous to the proof of Theorem 1. $\diamond$.

From the Theorem 2 it follows that
$\forall_{n,r,i,j} C^{k,n,r}_{i,j} = C_0$ and $C_0$ is given by (\ref{MaxAchievTransRateNoInter}).
Because the transmission time between $i$-th and $j$-th node, when the data is transmitted along the
edge $T^{k,n,r}_{i,j}$ must be the same then we have the following relation
$$\forall_{n,i,j} \; t^{k,n,r}_{i,j} = t^{k}_{r}.$$
The formula (\ref{qij-cij-tij}) written in terms of trees for each set $U^{k,n,r}_{i,j}$ has the form
$$ \forall_{(i,j)\in T^{k,r}} \;\; q^{k}_{r} = c^{k,n,r}_{i,j} t^{k,n,r}_{i,j}.$$
The maximum network lifetime problem of the broadcast data transmission service
without interference can be written in the form
\begin{equation} \label{DefinitionOfMLBTPForTreesSinr0}
\left\{   \begin{array}{l}
 \min_{t^{k}} \max_{i\in S_N} \{ E_{i}^{}(t^{k}) \}_{i=1}^{N}, \;k\in [1,N], \\
E_{i}(t^{k})= P_0 \sum_{j,r}^{} \gamma_{i,j}^{-1} \;t^{k}_{r} \; T^{k,r}_{i,j},\\
c^{}_{0} \sum_{i,r} t^{k}_{r} \;T^{k,r}_{i,j} = Q_k,\;\; j\in [1,N],\\
t^{k}_{r}\geq 0, \;0\leq c_0 \leq C_0, \;\;T^{k,r} \in V_{k}^{}, \; r\in [1,N^{N-2}].
\end{array} \right.
\end{equation}
%
The following lemma gives the solution of the problem (\ref{DefinitionOfMLBTPForTreesSinr0})
for the internal nodes of the one dimensional regular sensor network $L_N$.
%

{\bf Lemma 2}.
The solution of the maximum network lifetime problem
for the broadcast data transmission service (\ref{DefinitionOfMLBTPForTreesSinr0})
in the one dimensional regular sensor network $L_N$ with the channel gain function
(\ref{gamma-1[r]}) and channel capacity (\ref{MaxAchievTransRateNoInter})
is given by the following set of data transmissions trees $T^r$
\begin{equation} \label{OptimalTree2-N-Minus1-ver5}
   \begin{array}{ll}
 (T^{r}_{k,k-1},...,T^{r}_{2,1}, T^{r}_{r,k+1},T^{r}_{k+1,k+2},..., T^{r}_{N-1,N}),  &r\in [1,k-1],\\
 (T^{k}_{k,k-1},..., T^{k}_{2,1}, T^{k}_{k,k+1},..., T^{k}_{N-1,N}\},                  &r=k,\\
(T^{r}_{k,k+1}, ..., T^{r}_{N-1,N}, T^{r}_{r,k-1}, T^{r}_{k-1,k-2},..., T^{r}_{2,1}),  & r \in [k+1,N], \\
\end{array} \end{equation}
and weights $t^k_r$
\begin{equation} \label{t^k_i}  
t^k_r  = \left\{   \begin{array}{ll}
  \frac{\gamma^{-1}_1}{\gamma^{-1}_k} t^k_k + \frac{\gamma^{-1}_1}{\gamma^{-1}_k} \frac{1}{C_0}Q_k,  & r=1, \\
  \frac{\gamma^{-1}_1}{\gamma^{-1}_{k+1-r}} t^k_k,                                         & r\in [2, k-1], k\geq 3,\\
  \frac{\gamma^{-1}_1}{\gamma^{-1}_{r-k+1}} t^k_k,                                         & r\in [k+1, N-1], k\leq N-2, \\
  \frac{\gamma^{-1}_1}{\gamma^{-1}_{N-k+1}} t^k_k + \frac{\gamma^{-1}_1}{\gamma^{-1}_{N-k+1}} \frac{1}{C_0} Q_k, & r=N, \\
\end{array} \right.\end{equation}
where 
\begin{equation} \label{t^k_k}
t^k_k = \frac{1 - \frac{\gamma^{-1}_1}{\gamma^{-1}_k} - \frac{\gamma^{-1}_1}{\gamma^{-1}_{N-k+1}} }
             {-1+ \sum_{i=1}^{k} \frac{\gamma^{-1}_1}{\gamma^{-1}_i}+ \sum_{i=1}^{N-k+1} \frac{\gamma^{-1}_1}{\gamma^{-1}_{i}} } \;\frac{1}{C_0} Q_k.
\end{equation}
and $\gamma^{-1}_r$, $r\in [1,N]$ is given by (\ref{gamma-1[r]}).

Proof of this lemma can be found in \cite{Lipinski2}.
For the boundary nodes $i=1,N$ of $L_N$ the solution of (\ref{DefinitionOfMLBTPForTreesSinr0}) is trivial.
Proof of this fact can be found in \cite{Lipinski2}.
%
 \section{Gupta-Kumar model without interference}
We apply the results of Theorem 1 and Theorem  2 to the maximum network lifetime problem
with channel capacity function given by (\ref{SINR-Gupta-Kumar}).
If nodes of the network transmit data with the some constant power $P(x_i)=P_0$
and there is no interference,
then the maximum transmission rate in (\ref{SINR-Gupta-Kumar}) is given by the expression
\begin{equation} \label{MaxAchievTransRateNoInter-G-K}
C(x_i,x_j) = \log (1 + \frac{P_0}{\textit{N}_o} \gamma(x_i,x_j) ).
\end{equation}
From (\ref{MaxAchievTransRateNoInter-G-K}) it follows that even when there is no interference
the function $C(x_i,x_j)$ still depends on the distance between transmitter and receiver.
In models, in which the SINR function (\ref{SINR-Gupta-Kumar}) is utilized,
the nodes of the sensor network do not adjust the transmission power to the distance of the receiver.
Thus, we must modify the formulas for the energy of the node (\ref{NodeEnergyUnSINR}) and (\ref{NodeEnergySinrBroadcast}).
The maximum network lifetime problem of the multipoint-to-multipoint transmission service
with the channel capacity (\ref{MaxAchievTransRateNoInter-G-K}) has the form
 \begin{equation} \label{MLProblemM2M-NoInterferece-G-K}
 \left\{   \begin{array}{l}
   \min_{t} \max_{i} \{ E_{i}(t)\},  \\
E_i(t) = P_0 \sum_{j} t_{i,j},\\
\sum_{i} c_{i,j} t_{i,j} = Q_i + \sum_{j} c_{j,i} t_{j,i},\\
t_{i,j}\geq 0, \; 0 \leq c_{i,j} \leq C_{i,j}, \;\; i,j\in [1,N],
  \end{array} \right.\end{equation}
where $C_{i,j}\equiv C(x_i,x_j)$ is given by (\ref{MaxAchievTransRateNoInter-G-K}).
In (\ref{MLProblemM2M-NoInterferece-G-K}) we put $c_{i,j} = C_{i,j}$
and rescaled the time variables as follows
$t_{i,j} \rightarrow  \frac{C_0}{C_{i,j}}t_{i,j}$,
where $C_0$ is an arbitrary positive number of dimension $[b/s]$.
Because $\log(1+ x)\approx x$ for $|x|< 1$, then for
$\frac{P_0}{\textit{N}_o} \gamma(x_i,x_j) < 1$   
we can write
$$C_{i,j} = \log (1 + \frac{P_0}{\textit{N}_o} \gamma(x_i,x_j)) \approx \frac{P_0}{\textit{N}_o} \gamma(x_i,x_j).$$
This means, that the maximum lifetime problem for the multipoint-to-multipoint transmission service
can be transformed to the form (\ref{MLProblemM2PNoInterferece}), where
$E_i(t) = \textit{N}_o C_0 \sum_{j} \gamma^{-1}_{i,j}\; t_{i,j}$ and $c_0=C_0$.
%
The requirement $\frac{P_0}{\textit{N}_o} \gamma(x_i,x_j) < 1$, for
$\gamma_{i,j} = (d_{i,j})^{-a}$ can be written as
$\frac{P_0}{\textit{N}_o} < (d_{i,j})^{a}$,
which means that for a sufficiently large network
the two models (\ref{SINR-Gupta-Kumar}), (\ref{MaxTransRateG-K}) and (\ref{MaxAchievTransRateModel2}) coincide.

Similar results can be obtained for the broadcast data transmission service with the channel capacity function
(\ref{SINR-Gupta-Kumar}).
For the broadcast data transmission service the third formula in (\ref{MLProblemM2M-NoInterferece-G-K})
should be replaced by the third formula in (\ref{DefinitionOfMLBTPForTreesSinr0}) and the
energy of each node by $E_{i}(t^{k})= P_0 \sum_{j,r}^{} t^{k}_{r} \; T^{k,r}_{i,j}$.
\section{Conclusions}
In the paper we analyzed the maximum network lifetime problem
in sensor networks with limited channel capacity
for multipoint-to-multipoint and broadcast data transmission services.
We showed, that for the optimal data transmission of the maximum lifetime problem there is no interference of signals
in the network.
We introduced new Signal to Interference plus Noise Ratio function which was defined under the assumption that
the signal power $P(x_i)$ of the transmitting node depends on the signal gain $\gamma(x_i,x_j)$
and location of the receiver $P(x_i)=P_0 \gamma(x_i,x_j)$.
For such transmission model with the channel capacity defined by the new SINR function
we solved the maximum lifetime problem
in one dimensional regular sensor network $L_N$
for both data transmission services.
We showed, that these solutions coincide with the solutions of the maximum lifetime problem
analyzed in \cite{Giridhar, Chang, Lipinski1, Lipinski2},
where the problem was defined in terms of data transmission cost energy matrix $E_{i,j}$ and data
flow matrix $q_{i,j}$. This way we find the relation between the energy consumption model  written in terms of
$E_{i,j}$ and $q_{i,j}$ and the model defined in terms of signal power, transmission time and channels capacities.

There are several differences between the SINR functions $s(x_i,x_j,U^n_{i,j})$ defined in (\ref{MaxAchievTransRateModel2})
and proposed by Gupta and Kumar in (\ref{SINR-Gupta-Kumar}).
It seems that, the most interesting future of the proposed SINR functions 
is that for a gain function of the form $\gamma(x_k,x_m)=d(x_k,x_m)^{-a}$
the formula (\ref{MaxAchievTransRateModel2}) is invariant under
the rescaling of the distance between network nodes $d^{'}(x_k,x_m)=\lambda d(x_k,x_m)$.
From which follows, that the solutions of the maximum lifetime problem for two networks with proportional distance
between nodes are related by a constant factor.
For example, when we extend the distance between all nodes by the same factor $d(x_k,x_m) \rightarrow \lambda d(x_k,x_m)$,
and the node energy for the distance $d(x_k,x_m)$ is equal
$E_i(t) = P_0 \sum_{j,n} d_{i,j}^{a} \; t^n_{i,j}$,
then the nodes must increase their power signal and
the energy consumption increases to
$E_i(t) = \lambda^{a} P_0 \sum_{j,n} d^{a}_{i,j} \; t^{n}_{i,j}$.
\end{document}